\documentclass[submission, Phys]{SciPost}

\hypersetup{
    colorlinks,
    linkcolor={red!50!black},
    citecolor={blue!50!black},
    urlcolor={blue!80!black}
}

\usepackage[bitstream-charter]{mathdesign}
\DeclareSymbolFont{usualmathcal}{OMS}{cmsy}{m}{n}
\DeclareSymbolFontAlphabet{\mathcal}{usualmathcal}

\fancypagestyle{SPstyle}{
\fancyhf{}
\lhead{\colorbox{scipostblue}{\bf \color{white} ~SciPost Physics }}
\rhead{{\bf \color{scipostdeepblue} ~Submission }}

\fancyfoot[C]{\textbf{\thepage}}
}

\usepackage{amsthm}
\usepackage{amsmath}
\usepackage{mathtools}
\usepackage{graphicx}
\usepackage{bm}
\usepackage{color}
\usepackage{enumitem}
\usepackage{verbatim}
\usepackage{braket}
\usepackage{lipsum}

\usepackage{tikz}

\usepackage{graphicx}

\newcommand{\tr}{\text{Tr}}
\newcommand{\ave}[1]{\langle #1 \rangle}
\newcommand{\cl}[1]{\hat{\mathcal{#1}}}

\newcommand{\new}[1]{\textcolor{black}{#1}}

\begin{document}

\pagestyle{SPstyle}

\begin{center}{\Large \textbf{\color{scipostdeepblue}{
Generalized Gibbs ensembles in  weakly interacting dissipative systems and digital quantum computers\\
}}}\end{center}

\begin{center}\textbf{
Iris Ul\v{c}akar\textsuperscript{1, 2$\star$} and
Zala Lenar\v{c}i\v{c}\textsuperscript{1$\dagger$}
}\end{center}

\begin{center}
{\bf 1} Jo\v{z}ef Stefan Institute, 1000 Ljubljana, Slovenia
\\
{\bf 2} University of Ljubljana, Faculty for physics and mathematics, 1000 Ljubljana, Slovenia
\\[\baselineskip]
$\star$ \href{mailto:email1}{\small iris.ulcakar@ijs.si}\,,\quad
$\dagger$ \href{mailto:email2}{\small zala.lenarcic@ijs.si}
\end{center}

\section*{\color{scipostdeepblue}{Abstract}}
\textbf{\boldmath{%
Identifying use cases with superconducting circuits not critically affected by the inherent noise is a pertinent challenge. Here, we propose using a digital quantum computer to showcase the activation of integrable effects in weakly dissipative integrable systems. Dissipation is realized by coupling the system's qubits to ancillary ones that are periodically reset. We compare the digital reset protocol to the usual Lindblad \new{continuous-time} evolution by considering non-interacting integrable systems dynamics, which can be analyzed using scattering between the Bogoliubov quasiparticles caused by the dissipation. \new{If not dominant,} the inherent noise would cause extra scattering but would not critically change the physics. A corresponding quantum computer implementation would illuminate the possibilities of stabilizing exotic states in nearly integrable quantum materials.
}}

\vspace{\baselineskip}

\noindent\textcolor{white!90!black}{%
\fbox{\parbox{0.975\linewidth}{%
\textcolor{white!40!black}{\begin{tabular}{lr}%
  \begin{minipage}{0.6\textwidth}%
    {\small Copyright attribution to authors. \newline
    This work is a submission to SciPost Physics. \newline
    License information to appear upon publication. \newline
    Publication information to appear upon publication.}
  \end{minipage} & \begin{minipage}{0.4\textwidth}
    {\small Received Date \newline Accepted Date \newline Published Date}%
  \end{minipage}
\end{tabular}}
}}
}




\vspace{10pt}
\noindent\rule{\textwidth}{1pt}
\tableofcontents
\noindent\rule{\textwidth}{1pt}
\vspace{10pt}

\section{Introduction}
Most of the quantum simulators and computers strive to eliminate any elements of openness, however, to some extent, it is unavoidable: atom loss and dipolar coupling in cold atoms, light leakage in cavities, heating, dephasing and other errors on gates, etc. In the pioneering experiments with trapped ions  \cite{barreiro11, yu25} and also in some more recent experiments with superconducting qubit platform \cite{mi23}, there have been propositions on how to actually use engineered dissipation \cite{verstraete09,harrington22} to prepare target/ground states \cite{kraus08,chen23} or to measure phase transitions \cite{fang24}. Such protocols might also be more resilient to the inherent platforms' noise. For example, in a recent implementation of Trotterized transverse field Ising model with the superconducting circuit \cite{mi23}, a dissipative cooling towards the ground state has been implemented by coupling the system's qubits to ancilla ones that are periodically reset. This realization builds on a series of theoretical works \cite{matthies22,kaplan17,wang17,feng22,polla21,zaletel21,metcalf20,kishony23} proposing cooling in quantum computers by coupling to low entropy baths (ancilla qubits), involving tuning the Hamiltonian of the ancilla qubits and its coupling to the system qubits.
While the above mentioned cooling protocols might be more naturally and efficiently implemented with an ergodic system \cite{matthies22}, considering non-interacting models can assist to get more exact/analytical insight into the conditions required \cite{lloyd2024}. 

In many cases, non-interacting many-body models are the cornerstone of our understanding and description of many-body physics. The fact that they are exactly diagonalizable via the Bogoliubov transformation makes them also a rare and appealing platform to study non-equilibrium many-body physics \cite{essler16,vidmar16,dalessio016}. 
In the context of thermalization or its failure, non-interacting systems are an example of models with extensively many conserved quantities \cite{essler16,vidmar16}. The conserved quantities of translationally invariant models are simply the mode occupation operators of Bogoliubov quasiparticles \cite{vidmar16} and one can use those to construct extensively many local conserved quantities \cite{essler16}. The existence of macroscopically many conserved quantities places non-interacting many-body systems on the same footing as more general interacting integrable systems, in the sense that they fail to thermalize due to the presence of additional conservation laws, or equivalently, limited quasiparticle scattering \cite{vidmar16,essler16,dalessio016}. Non-interacting models have been among the first for which the applicability of generalized Gibbs ensembles (GGEs) \cite{rigol07} as a local description of steady states reached after a sudden quench has been demonstrated \cite{calabrese11gge,calabrese12agge,calabrese12gge,essler12,fagotti13,fagotti13gge2,bucciantini14}. Introducing additional Lagrange parameters, associated with the mode occupation operators or the local conserved operators, proved to be a successful way to take into account constraints on equilibration. More recent studies showed that a GGE description applies not only to quenches in isolated models but also to weakly dissipative integrable systems, including the non-interacting ones \new{\cite{lange17,lenarcic18,lange18,reiter21,schmitt22,ulcakar23,bouchoule20,rossini21,mazza22,mazza24,riggio24,gerbino23,perfetto23,rowlands23,lehr25,starchl22,starchl24}}. In that case, GGE \new{is expected to give} the zeroth order approximation to the dynamics and the steady state density matrix. \new{Such GGE descriptions have been largely used in the context of reaction-diffusion models \cite{gerbino23,perfetto23,rowlands23,lehr25} and particle/atom loss in quantum simulators\cite{bouchoule20,rossini21,mazza22,mazza24,riggio24}.} The main difference between the closed and open setup is that for the former, the Lagrange multipliers are determined by the post-quench state, while in the open setup, they are determined by the dissipation operator \new{which can stabilize a nontrivial steady state} \cite{lange17,lenarcic18,lange18,reiter21,schmitt22,ulcakar23,perfetto23,rowlands23,lehr25}. 
Only if the dissipator obeys the detailed balance condition, the stabilized steady state is thermal \cite{lenarcic18,lloyd2024}. In any other situations, such weakly dissipative, nearly integrable systems tend to converge to highly non-thermal GGEs. This explains why a careful tuning of parameters and coupling operators is necessary for an approximate ground state preparation on a quantum computer simulating an integrable system \cite{mi23,lloyd2024}. 

In this work, we marry the two topics and show that for generic weak couplings between integrable system and periodically reset ancilla qubits, highly non-thermal generalized Gibbs ensembles would be stabilized with quantum computers. We focus on the non-interacting integrable systems, for which we also review and compare different approaches to thermodynamically large systems. 
In Sec.~\ref{sec::approach}, we review the general description of weakly open integrable systems in terms of time dependent generalized Gibbs ensembles.
For non-interacting integrable models, this is reformulated in Sec.~\ref{sec::model} as a generalized scattering theory between the Bogoliubov quasiparticles for a weakly dissipative \new{continuous-time} Lindblad model with transverse field Ising model coupled to Lindblad baths.
 \new{We compare the time evolution of the scattering theory to a tensor network simulation and observe coincidence in the weak coupling limit, numerically showing that GGEs are indeed the correct zeroth order description of dynamics.}
In Sec.~\ref{sec::SCQubits} we highlight that superconducting circuit platforms \cite{mi23} or digital trapped ion quantum computers \cite{ringbauer22} would be an ideal implementations of all elements required to show that highly non-thermal and possibly exotic GGEs emerge in weakly open nearly integrable systems. To make the connection, we derive the effective equation of motion \new{for the system's density matrix} for the Floquet time propagated coupled system and ancilla qubits, involving the reset of ancilla qubits to implement the openness. This can be once again recast as a generalized scattering theory between the Bogoliubov quasiparticles.
In the end, we propose how reviving of integrability can be detected via measurement of anomalously slow decay of certain spatial correlations. 
In Sec.~\ref{sec::conclusions}, we conclude that an actual experimental realization would prove the concept of GGEs to be applicable also for other platforms and, ultimately, for nearly integrable materials \cite{scheie21,scheie22}.

\section{Setup}\label{sec::approach}
We first review the structure of the density matrix perturbation theory using the example of a traditional Lindblad setup with a \new{continuous-time} model \new{\cite{lange17,lenarcic18,lange18,reiter21,schmitt22,ulcakar23,bouchoule20,rossini21,gerbino23,perfetto23,rowlands23,lehr25,riggio24,starchl22,starchl24}}.  In Sec.~\ref{sec::SCQubits}, we generalize this to a Trotterized implementation with a reset protocol, relevant to digital quantum computers. 
Within the \new{continuous-time} implementations, we consider a system with dominant unitary dynamics described by a non-interacting, \new{one-dimension} translationally invariant Hamiltonian $H_0$, which has a diagonal form in terms of mode occupation operators $n_q$ of Bogoliubov quasiparticles, 
\begin{equation}
H_0 = \sum_q \varepsilon_q n_q + E_0
\end{equation} 
where $\varepsilon_q$ is the dispersion of a single particle excitation with momentum $q$ and $E_0$ is a constant shift in energy. In addition, the system is weakly coupled in \new{the} bulk to baths described by the dissipator $\cl{D}$,
\begin{equation}
\cl{L} \rho = -i [H_0,\rho] + \cl{D} \rho, \ 
 \cl{D} \rho = \epsilon \sum_i L_i \rho L_i^\dagger - \frac{1}{2}\{L_i^\dagger L_i, \rho\}.
 \label{eq::liouvillian}
\end{equation}
Here, $\epsilon\ll 1$ is a weak coupling parameter, and $L_i$ are the Lindblad operators acting \new{at or in a small region} around site $i$.

In our previous works \cite{lange17,lange18,reiter21,lenarcic18}, we \new{discussed} that the zeroth order (in $\epsilon$) approximation to the steady state and the slow evolution towards the steady state has the form of a generalized Gibbs ensemble (GGE). For the non-interacting \new{one-dimensional} translationally invariant $H_0$ one can build a GGE using the local extensive conserved quantities $C_i$, $[H_0, C_i]=0$, or the mode occupation operators $n_q$ \new{\cite{lange17,lenarcic18,lange18,reiter21,schmitt22,ulcakar23,bouchoule20,rossini21,gerbino23,perfetto23,rowlands23,lehr25,riggio24,starchl22,starchl24}}, 
\begin{equation}\label{eq::GGE}
\rho_{\boldsymbol{\mu}}(t) = \frac{e^{-\sum_q \mu_q(t) n_q}}{\tr[e^{-\sum_q \mu_q(t) n_q}]}.
\end{equation}
Here, $\mu_q$ are the associated Lagrange multipliers. Since the dissipator weakly breaks the integrability properties of $H_0$, mode occupations are slowly changing, in the lowest order described by the rate equations 
\begin{equation}\label{eq::nkdot}
\ave{\dot{n}_{q}}(t) 
\approx \tr\left[n_{q} \cl{D} \frac{e^{-\sum_{q'} \mu_{q'}(t) n_{q'}}}{\tr[e^{-\sum_{q'} \mu_{q'}(t) n_{q'}}]}\right],
\end{equation}
where contribution of order $\epsilon^2$ and higher are neglected. Equivalently, the Langrange multipliers $\mu_q$ will be changing on the timescale $\mathcal{O}(1/\epsilon)$ according to the following equation derived in Ref.~\cite{lange18},
\begin{equation}\label{eq::mudot}
\dot{\mu}_q(t) = - \chi^{-1}_{q,q}(t) \, \ave{\dot{n}_{q}}(t), \quad  \chi_{q,q}(t)= \frac{e^{-\mu_q(t)}}{(1+e^{-\mu_q(t)})^2}.
\end{equation}
Here, we used $\ave{O}(t) \equiv \tr[O\rho_{\boldsymbol{\mu}}(t)]$ and that $\chi$ matrix with
$\chi_{q,q'}(t) = \ave{n_q n_{q'}} - \ave{n_{q}}\ave{n_{q'}}$
entries is diagonal for free fermions.

\new{If the time evolution given by Eq.~\eqref{eq::mudot} is performed up to long times $\mathcal{O}(1/\epsilon)$, we eventually converge to the steady state.} We should note that this is only one possible approach to the steady state calculation. In App.~\ref{sec::approaches}, we review alternative direct  steady state calculations where Lagrange multipliers are determined via the root finding procedure for the stationarity condition for: (i)~all mode occupations $n_q$, Eq.~\eqref{eq::nkdot},  (ii) iteratively constructed leading conserved quantities \cite{ulcakar23}, and (iii) for the most local conserved quantities $C_i$ \cite{lange17}. In App.~\ref{sec::approaches}, we also compare the scaling complexity of those different approaches.

\section{Continuous-time Model}\label{sec::model}
We consider the transverse field Ising model 
\begin{equation}\label{eq:H0spin}
H_0 = \sum_i J \sigma^x_i \sigma ^x_{i+1} + h \sigma ^z_i,    
\end{equation}
as a paradigmatic non-interacting integrable model, which can be (at least approximately) realized with quantum simulators \cite{kim11,islam11,labuhn16,zhang17,Ebadi2020,Scholl2020}. In order to obtain its mode occupation operators, we perform the Jordan-Wigner tranformation from spin-$\frac{1}{2}$ degrees of freedom to spinless fermions
\begin{align}\label{eq::JW}
    \sigma_j^z &= 2c_j^{\dagger}c_j - 1,\
    \sigma_j^{+} = e^{i\pi\sum_{l<j}n_l}c_j^{\dagger},
\end{align}
and the Fourier transform from the positional basis to the momentum basis
\begin{align} \label{eq::FTfermion}
    c_j &= \frac{e^{-i\pi/4}}{\sqrt{L}} \sum_q e^{iqj} c_q.
\end{align}
Finally, the Bogoliubov transformation
\begin{align}\label{eq::disp}
    c_q = u_q d_q - v_qd_{-q}^{\dagger},
\end{align}
\new{parametrized with
\begin{align}\label{eq::disp-1}
    u_q = \frac{\varepsilon_q+a_q}{\sqrt{2\varepsilon_q(\varepsilon_q + a_q)}}, \,
    v_q = \frac{b_q}{\sqrt{2\varepsilon_q(\varepsilon_q + a_q)}}, \,
    a_q = 2(J\cos{q}+h),\,
    b_q = -2J\sin{q},
\end{align}
}
brings the Hamiltonian into a diagonal form
\begin{equation}\label{eq::disp-2}
    H = \sum_q \varepsilon_q \left(n_q -\frac{1}{2}\right), \ \varepsilon_q = 2\sqrt{J^2 + 2 h J \cos{q} + h^2}, \ n_q=d_q^{\dagger}d_q.
\end{equation}
Therefore, the Hamiltonian and all the local conserved charges, $C_i = \sum_q c^{(i)}_q n_q$, can be expressed in terms of mode occupation operators $n_q$.
One should note that periodic boundary conditions in the spin picture are translated to periodic boundary conditions in the fermion picture for an odd number of particles and anti-periodic for an even number of particles. Consequently, the two cases are diagonalized by a different set of wave vectors, $\mathcal{K}^{+}=\{\frac{2\pi}{L}(q+\frac{1}{2}), q = 0, \dots L-1\}$ for the even sector and $\mathcal{K}^{-}=\{\frac{2\pi}{L}q, q = 0, \dots L-1\}$ for the odd sector. The two symmetry sectors are uncoupled by the Hamiltonian dynamics and should be treated separately.

As an example of coupling to baths that stabilize a nontrivial steady state we consider the following Lindblad operator
\begin{equation}
    L_j = S_j^{+}S_{j+1}^{-} + S_j^{z} + \frac{1}{2}\mathbb{1}_j.
\end{equation}
We choose an operator which after the Jordan-Wigner, Fourier and Bogoliubov transformations obtains a compact form without any string operators,
\begin{equation}\label{eq::Li}
    L_j =  \sum_{q, q'} \frac{e^{-ij(q-q')}}{L}(1+e^{iq'}) (u_qd_q^{\dagger} - v_qd_{-q})(u_{q'}d_{q'} - v_{q'}d_{-q'}^{\dagger}).
\end{equation}
However, due to the form of dissipator with $L_i$ and $L_i^\dagger$ pairs, Eq.~\eqref{eq::liouvillian}, analysis is not much more complicated in the presence of string operators as well.
These Lindblad operators preserve the parity \new{of the number of fermions}, i.e., some terms preserve the number of fermions while others change it by two. Therefore, we get two steady states, one for the even and one for the odd parity sector. Thermodynamically, the two solutions behave the same. We consider only the even sector in the following and work with momenta $\mathcal{K}^{+}$.

\begin{figure}[t!]
\includegraphics[width=01\columnwidth]{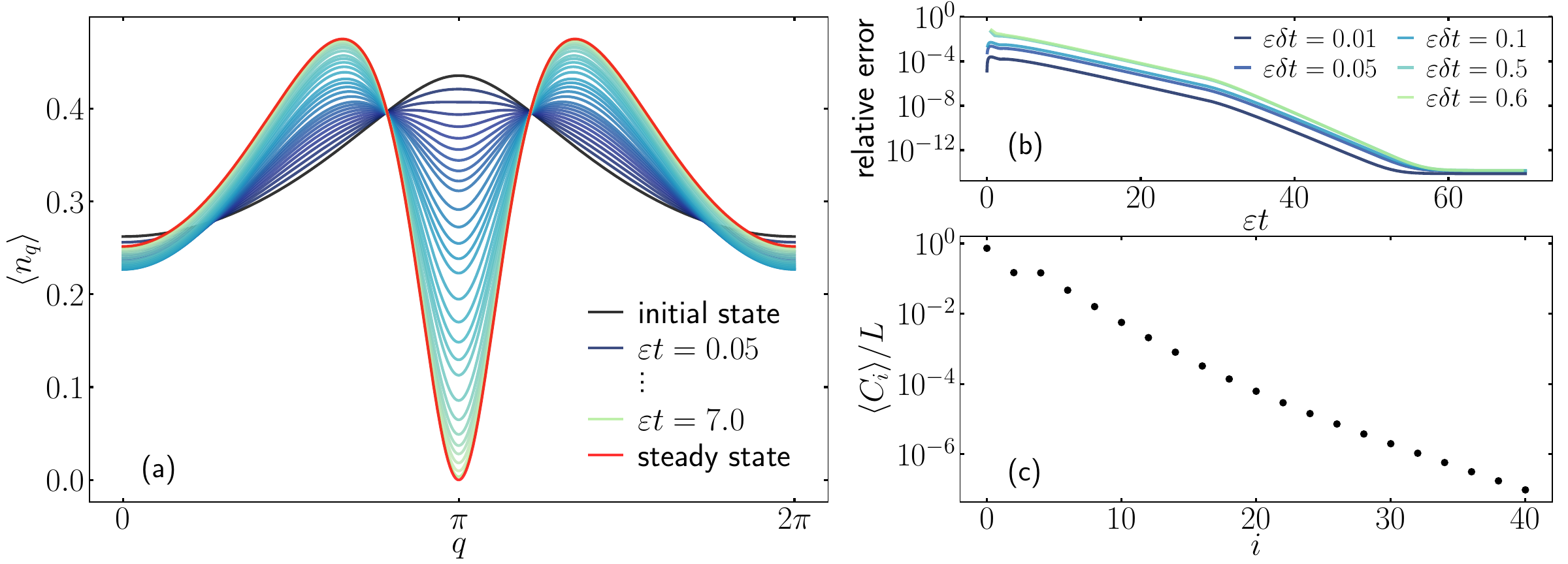}
\caption{\new{(a)} Time evolution from an initial thermal mode occupation with $\beta=0.323$ to a highly non-thermal steady state distribution, stabilized by our choice of Lindblad operators, Eq.~\eqref{eq::Li}. 
(b) Relative error 
$\sum_q |(\langle n_q\rangle (t) - \langle n_q\rangle_0 (t))
/ \langle n_q\rangle_0(t) |/L$ of the occupations $\langle n_q\rangle (t)$ obtained with Euler method with time steps $\epsilon\delta t=0.01,...,0.6$ and the reference $\langle n_q\rangle_0 (t)$ evaluated with smallest $\epsilon\delta t= 0.005$.
At late times differences are tiny. (c) Steady state expectation values of local conserved quantities \eqref{eq::isingC}. With increasing support, the importance of even conserved quantities decays exponentially. Expectation values of odd observables are zero due to symmetry.
Parameters: $J = 1, h = 0.6, L=10^5$.}
\label{fig1}
\end{figure}

To calculate the time evolution as described in Sec.~\ref{sec::approach}, the central object to be evaluated is the expression  \eqref{eq::nkdot} for $\ave{\dot{n}_q}$, which can be split as
\begin{equation}
    \label{eq:terms}
    \ave{\dot{n}_q} = \epsilon \sum_j \langle L_j^{\dagger}n_q L_j\rangle -\langle n_q L_j^{\dagger}L_j\rangle 
\end{equation}
Here, we took into account the cyclicity of trace and the expectation value $\ave{\cdot}$ with respect to the GGE $\rho_{\boldsymbol{\mu}}(t)$, Eq.~\eqref{eq::GGE}. 
Due to the diagonal form of the GGE, only the combinations of creation $d^\dagger_{q}$ and annihilation $d_{q}$ operators, which are in total diagonal in the mode occupation operators, contribute to the expectation values with respect to the GGE Ansatz. 
After extracting the contributing Wick contractions and simplifying the remaining terms, Eq.~\eqref{eq:terms} obtains a compact and meaningful form
\new{\begin{align}\label{eq::nqdot}
\langle \dot{n}_q \rangle = &\frac{2\epsilon}{L} \sum_{q'}  
 f^s_{q, q'} \langle 1 - n_q \rangle \langle n_{q'} \rangle - f^s_{q', q} \langle n_q \rangle \langle 1 - n_{q'} \rangle + f^c_{q, q'} \langle 1 - n_q \rangle \langle 1 - n_{q'} \rangle - f^a_{q, q'} \langle n_q \rangle \langle n_{q'} \rangle. 
\end{align}}
The first two terms correspond to the transitions between $q'$ and $q$ momenta, weighted by a parameter-dependent positive function 
\new{\begin{align}\label{eq::f1q}
f^s_{q, q'} = &u_q^2 u_{q'}^2 (1+\cos q') + v_q^2 v_{q'}^2(1+\cos q)- u_q v_q u_{q'} v_{q'} (1+\cos{q'} +\cos q +\cos(q+q') ),
\end{align}}
while the last two terms correspond to creation/annihilation of $q'$ and $q$ modes, weighted by \new{positive functions
\begin{align}\label{eq::f2q}
f^c_{q', q} = &v_q^2 u_{q'}^2 (1+\cos {q}) + u_q^2 v_{q'}^2(1+\cos q')- u_q v_q u_{q'} v_{q'} (1+\cos {q'} +\cos q +\cos(q-{q'}) ),\notag\\
f^a_{q', q} = &v_q^2 u_{q'}^2 (1+\cos {q'}) + u_q^2 v_{q'}^2(1+\cos q)- u_q v_q u_{q'} v_{q'} (1+\cos {q'} +\cos q +\cos(q-{q'}) ).
\end{align} }
Terms with $\ave{1-n_q}$, corresponding to transitions into the $q$ mode, have a positive sign. On the other hand, terms with $\ave{n_q}$, where $q$ mode is annihilated, have a negative sign. 
In the GGE, the expectation value of the mode occupation operator is given by $\ave{n_q} = e^{-\mu_q}/(1+e^{-\mu_q}).$
The rate equation \eqref{eq::nqdot} thus has the structure of the Boltzmann equation but without the usual assumption of thermal Fermi functions. 

We should note that \new{$f^s_{q,q'}, f^c_{q,q'}$ and $f^a_{q,q'}$} can be factorized over variables $q,q'$ and therefore summation over $q'$ in  Eq.~\eqref{eq::nqdot} can be performed independent of $q$. The complexity of evaluating $\ave{\dot{n}_q}$ for all $q$ thus scales as $\mathcal{O}(L)$. A similar factorization property over an arbitrary number of momentum varibles should also hold for other choices of Lindblad operators, implying that $\ave{\dot{n}_q}$ is calculated in $\mathcal{O}(L)$ generically. 

We perform calculations of time-dependent Lagrange parameters $\mu_q(t)$ from Eq.~\eqref{eq::mudot} by summation over discrete momenta on $L=10^5$ sites. Fig.~\ref{fig1}(a) shows how the momentum distributions change from an initial thermal distribution around $q=\pi$ (where the minimum of dispersion $\varepsilon_q$, Eq.~\eqref{eq::disp-2}, lies for our choice $J=1$, $h=0.6$), to a highly non-thermal distribution, double-peaked around some non-trivial momenta. This result is the main message of our example: since our Lindblad operators $L_i$, Eq.~\eqref{eq::Li}, do not obey detailed balance, a highly non-thermal steady state is stabilized even if the coupling to the baths is only weak.

The calculation is performed using the Euler method with time step $\epsilon\delta t =0.6$, which is sufficiently small that errors do not affect the dynamics significantly and the system converges to the right steady state. 
Namely, Fig.~\ref{fig1}(b) shows the difference between calculations done at chosen $\epsilon \delta t=[0.01,0.05,0.1,0.5,0.6]$ with respect to the smallest $\epsilon \delta t=0.005$ time step tested. 
In an absolute sense, the relaxation time is given by the strength of the coupling to the bath. i.e., the distributions relax to the steady state on $1/\epsilon$ timescale since the rate of change for the mode occupations is proportional to $\epsilon$, Eqs.~(\ref{eq:terms}, \ref{eq::nqdot}). However, for the same reason, we can use scaled $\epsilon \delta t$ in our discrete-time propagation scheme.
In App. \ref{sec::approaches}, we compare the performance of steady-state calculation from either direct time evolution used above or the iterative steady state construction introduced in Ref.~\cite{ulcakar23}. For calculations in the basis of mode occupation operators, the two approaches are comparable in the studied case.


Structured distribution of quasiparticle mode occupations, Fig.~\ref{fig1}(a), in the spin language implies a non-thermal steady state expectation values of local conserved quantities, $C_{2\ell} = \sum_{q} \cos{(q\ell)} \varepsilon_{q} n_q$ and $C_{2\ell-1} = 2J\sum_{q} \sin{(q\ell)} n_q$ \cite{fagotti13gge2}. Since the stabilized distribution is symmetric under momentum inversion $\ave{n_q} = \ave{n_{-q}}$, odd conserved quantities are not stabilized $\ave{C_{2\ell-1}}=0$. Fig.~\ref{fig1}(c) shows that the expectation values of even conserved quantities decay exponentially with their support, implying that a truncated GGE description involving the most local conserved quantities can be a reasonable approximation as well.

\begin{figure}[t!]
\includegraphics[width=1\columnwidth]{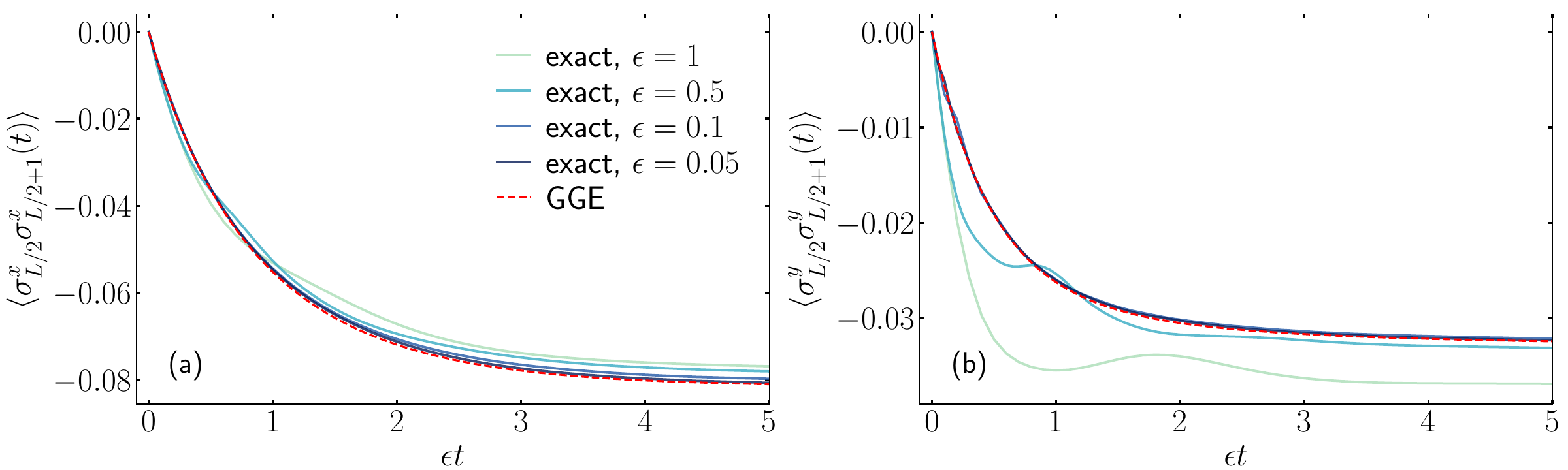}
\caption{\new{Comparison of exact and GGE Ansatz dynamics at scaled time $\epsilon t$ for (a) $\ave{\sigma^x_{L/2} \sigma^x_{L/2+1}}$ and (b) $\ave{\sigma^y_{L/2} \sigma^y_{L/2+1}}$. Exact evolution is obtained with tensor network representation for the full Liouvillian at four different coupling strengths of the dissipator $\epsilon = 1, 0.5, 0.1, 0.05$.
Parameters: $J = 1, h = 0.6, L=80, \chi=100$.}}
\label{fig2a}
\end{figure}

\new{While GGE Ansatz for a weakly open setup is very well physically motivated and used in several publications, it is an ansatz after all. Therefore we compare in Fig.~\ref{fig2a} the dynamics of $\ave{\sigma^x_{L/2} \sigma^x_{L/2+1}}$ and $\ave{\sigma^y_{L/2} \sigma^y_{L/2+1}}$, calculated using the GGE ansatz, Eq.~\eqref{eq::nqdot}, or with tensor network representation of the vectorized density matrix, evolved using time-evolving block decimation (TEBD) \cite{zwolak04}, for the full Liouvillian, Eq.~\eqref{eq::liouvillian}, at four different coupling strengths of the dissipator $\epsilon = 1, 0.5, 0.1, 0.05$. A sufficiently large bond dimension $\chi=100$ is used \new{and time evolution is performed from an infinite temperature initial state}. As coupling strength is reduced and the time-axis is rescaled accordingly, the exact results at finite couplings approach the GGE Ansatz solution. This numerically confirms that the GGE Ansatz is indeed the zeroth-order (in the coupling strength $\epsilon$) approximation to the exact density matrix, see also \cite{lumia24}. In order to reach larger system sizes, we used open boundary conditions for the tensor network simulations. Since our choice of Lindblads does not induce currents, $\ave{C_{2\ell-1}}=0$, a good comparison to the GGE results calculated with periodic boundary conditions (everywhere in the paper) is obtained for observables in the bulk.}

\new{We should note that a similar simplification of evolution using Wick's contractions has been used before to treat reaction-diffusion processes with two-particle loss as a source of dissipation \cite{gerbino23,perfetto23,rowlands23,lehr25,riggio24}. A related scattering theory has also been developed for non-interacting \cite{rossini21} and more involved interacting Lieb-Liniger gas in the presence of atom losses \cite{bouchoule20}. However, an important feature of our example is that our steady state is non-trivial and structured, highlighting the possibility of engineering highly non-thermal states via the interplay of integrability and dissipation.}

\section{Digital quantum computer protocol}\label{sec::SCQubits}
\begin{figure}[t!]
\centerline{\includegraphics[width=0.5\columnwidth]{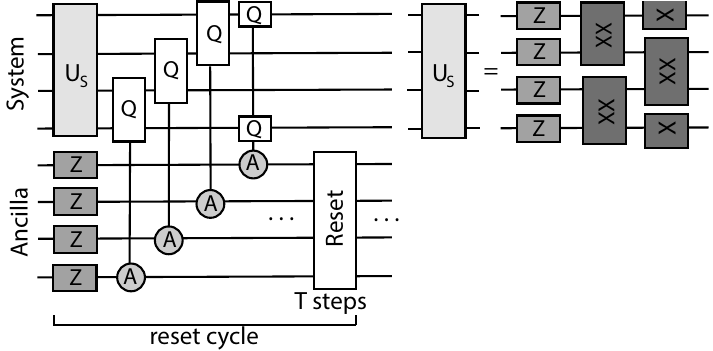}}
\caption{Scheme of our dissipative transverse field Ising realization, similar to Refs.~\cite{mi23,matthies22} and realistic to implement with a digital quantum computer. In this setup, the system's qubits are coupled to ancillary ones. After every $T$ system-ancilla-coupling propagations, ancilla qubits are reset to the spin-down state.}
\label{figSC}
\end{figure}

We continue by discussing a contemporary possible realization of such non-thermal states using a digital quantum computer. There, dissipation can be realized by coupling system's qubits to auxiliary ones and resetting the latter to, e.g., spin down state every $T$ steps~\cite{mi23}. A sketch of a possible realization is shown in Fig.~\ref{figSC}. While Ref.~\cite{mi23} used such a reset protocol for an approximate ground state preparation by dissipative cooling for the transverse field Ising model, we would like to point out that due to the proximity to integrability such a weakly dissipative setup is prone to realize highly non-thermal GGEs, with the steady state mode occupations fixed by the form of coupling to the ancilla qubits.

As the integrable system we again consider a transverse field Ising model, now realized via Trotterized gate propagation with gate duration chosen to be $\pi/2$,
\begin{equation}\label{eq::US}
U_S = e^{-i\frac{\pi J}{2}\sum_{j}\sigma^x_j\sigma^x_{j+1}} e^{-i\frac{\pi h}{2}\sum_{j}\sigma^z_j}
\equiv e^{-i H_{\text{FTFI}}},
\end{equation}
where $ H_{\text{FTFI}}$ is the corresponding Floquet Hamiltonian derived below. Ancilla qubits are propagated by simple
\begin{equation}\label{eq::UA}
U_A = e^{-i\frac{\pi h_A}{2}\sum_{j}\tilde\sigma^z_j},
\end{equation}
where $\tilde\sigma_j^\alpha$ represent operators acting on ancilla qubits. In addition, at each time step $\tau \le T$ within the reset cycle before the reset,
system and ancilla qubits are coupled by
\begin{equation}\label{eq::coupling1}
U_{SA, \tau} = \prod_j e^{ -i \lambda_{\tau} Q_j \otimes A_j}. 
\end{equation}
We use coupling operators resembling the Lindblad operators \eqref{eq::Li} from the previous section,
\begin{equation}\label{eq::coupling}
Q_j = S_j^{+}S_{j+1}^{-} + S_j^{-}S_{j+1}^{+}, \quad
A_j = \tilde\sigma_j^x,
\end{equation}
where $Q_j$ operators act on the system's qubits, while $A_j$ operators act on the ancilla qubits. Applying multi-qubit gates has been relized before \cite{satzinger21}. One cycle contains $T$ system-ancilla-coupling propagations
\begin{equation}\label{eq::totalU}
U_T=U_{SA, T} U_A U_S \cdots U_{SA, 1} U_A U_S, 
\end{equation}
followed by the reset of ancilla qubits to the down spin state. 

Following Ref.~\cite{lidar2019} and assuming that the coupling between the system and ancilla qubits is small $\lambda_\tau \ll 1$, we derive the system's density matrix interaction-picture evolution for one reset cycle, from cycle number $N_c$ to $N_c + 1$,
\begin{align}\label{eq::Tprop}
&\rho_{S,I}(N_c+1) - \rho_{S,I}(N_c) \\
&\hspace{0.3cm}\approx\sum_{j,\omega, \omega'} 
-i \, \rm{Im}(\mathcal{A}_{\omega,\omega'}) [Q_{j,\omega'}^\dagger Q_{j,\omega}, \rho_{\boldsymbol{\mu}}(N_c)]  + a_{\omega,\omega'}\Big(Q_{j,\omega} \rho_{\boldsymbol{\mu}}(N_c) Q_{j,\omega'}^\dagger - \frac{1}{2} \{Q_{j,\omega'}^\dagger Q_{j,\omega}, \rho_{\boldsymbol{\mu}}(N_c)\}\Big).\notag
\end{align}
Above we introduced ancilla correlation functions
\begin{align}
\label{eq::formfun}
\mathcal{A}_{\omega,\omega'}
&= \sum_{\tau=1}^{T}\sum_{\tau'=1}^{\tau} \lambda_\tau \lambda_{\tau'} e^{i(\omega'\tau - \omega \tau' + \pi h_A(-\tau+\tau'))}, \\
a_{\omega,\omega'} 
&= \sum_{\tau=1}^{T} \lambda_\tau e^{i\tau(\omega'-\pi h_A)} \sum_{\tau'=1}^{T} \lambda_{\tau'} e^{-i\tau'(\omega-\pi h_A)}. \notag
\end{align}
Operator $Q_{j,\omega} = \sum_{\alpha, \beta,\\ \tilde{E}_\beta - \tilde{E}_\alpha = \omega} \ket{\alpha}\bra{\alpha} Q_j \ket{\beta}\bra{\beta}$ represents $Q_j$, Eq.~\eqref{eq::coupling}, projected between the many-body eigenstates of the system's unitary operator $U_S$ that differ in quasi-energy for $\omega$. Namely, $\ket{\alpha}$ is a many-body eigenstate of the systems's unitary $U_S$ with a corresponding eigenvalue $e^{-i\tilde{E}_{\alpha}}$, where $\tilde{E}_{\alpha}$ is the \new{many-body} quasi-energy of the Floquet Hamiltonian $H_\text{FTFI}$. While operators $Q_j$ can be arbitrary, the form of ancilla correlation functions Eq.~\eqref{eq::formfun} is obtained from the specific choice of ancilla dynamics $U_A$, Eq.~\eqref{eq::UA}, and the coupling operator acting on the ancilla qubits $A_j$, Eq.~\eqref{eq::coupling}. Notably, the equation of motion \eqref{eq::Tprop} for the system's density matrix is of a Lindblad form. A general system's density matrix time evolution as well as more detailed derivation for our model are given in App.~\ref{app::rhoTEvo}.


We again use periodic boundary conditions for the system's gates under which the system's propagation operator factorizes over momenta
$
U_S = \prod_{q \geq 0} e^{-i \Phi_q^\dagger X_q \Phi_q} e^{-i \Phi^\dagger_q Z_q \Phi_q},
$
with $\Phi_q = \{c_q, c_{-q}^\dagger \}^T$ representing the bispinor of fermionic operators in momentum space, Eq.~\eqref{eq::FTfermion}. $X_q$ and $Z_q$ are 2x2 matrices, derived by representing the first and the second term in $U_S$, Eq.~\eqref{eq::US}, with fermionic operators in the momentum space, using relations (\ref{eq::JW}, \ref{eq::FTfermion}). Explicit expressions for $X_q, Z_q$ are given in App.~\ref{app::FTFI}, where we also derive that the Floquet quasi-energy dispersion $\tilde{\varepsilon}_q$ takes the form 
\begin{equation}
\cos(\tilde{\varepsilon}_q) = \cos(\pi J) \cos(\pi h) - \sin(\pi J) \sin(\pi h) \cos(q). 
\end{equation}
Coefficients $\tilde{u}_q, \tilde{v}_q$, connecting fermionic operators to the Bogoliubov ones, $c_q = \tilde{u}_q d_q - \tilde{v}^*_q d_{-q}^{\dagger}$, are for the Trotterized transverse field Ising model of the form 
\begin{align}
   \tilde u_q &= \frac{\xi_q+ \tilde a_q}{\sqrt{2\xi_q(\xi_q + \tilde a_q)}}, \
    \tilde v_q = \frac{\tilde b_q}{\sqrt{2\xi_q(\xi_q + \tilde a_q)}}, \
    \xi_q = \sqrt{\tilde a_q^2 + |\tilde b_q}|^2,\\
    \tilde a_q &= \sin(\pi J) \cos(\pi h) \cos(q) + \cos(\pi J) \sin(\pi h), \
    \tilde b_q = -e^{-i \pi h} \sin(\pi J) \sin(q), \notag
\end{align}
which is very similar to the original \eqref{eq::disp}. See App.~\ref{app::FTFI} for the derivation. Finally,
\begin{equation}
    H_{\text{FTFI}} = \sum_q \tilde{\varepsilon}_q \left(n_q-\frac{1}{2}\right), \ n_q=d_q^{\dagger}d_q.
\end{equation}
After the above mapping, the coupling operators $Q_j$ acting on the system qubits, Eq.~\eqref{eq::coupling}, obtain a bilinear form
\begin{align}\label{eq:Qj_ff}
    Q_j = \frac{1}{L} \sum_{q, q'} &e^{-ij(q - q')} (e^{-iq} + e^{iq'})
    (u_q d_q^{\dagger} - v_q d_{-q})(u_{q'} d_{q'} - v_{q'}^* d_{-q'}^{\dagger}).
\end{align}

In the case of weak coupling to ancilla qubits, $\lambda_\tau\ll 1$, changes within one reset cycle are small. Therefore, one can still use the Euler propagation method to calculate the time-dependent Lagrange multipliers, parametrizing $\rho_{\boldsymbol{\mu}}(N_c)$, from the rate equations for the $H_{\text{FTFI}}$ mode occupation operators. The latter obtains a compact and meaningful form, similar to the \new{continuous-time} model,
\new{\begin{align} \label{eq::nqdotGoogle}
&\ave{n_q(N_c+1)}-\ave{n_q(N_c)}
= \frac{2}{L} \sum_{q'} 
g^s_{q,q'} \big(\langle 1 - n_q \rangle \langle n_{q'} \rangle a_{\varepsilon_{q'}-\varepsilon_q} 
- \langle n_q \rangle \langle 1 - n_{q'} \rangle a_{\varepsilon_q-\varepsilon_{q'}}\big)   \\
&\hspace{5.3cm}+g^{ca}_{q,q'}\big(\langle 1 - n_{q} \rangle \langle 1 - n_{q'} \rangle a_{-\varepsilon_{q'}-\varepsilon_q} \notag - \langle n_q \rangle \langle n_{q'} \rangle a_{\varepsilon_{q'}+\varepsilon_q}\big). \notag
\end{align}}
For a GGE form of the density matrix, Eq.~\eqref{eq::Tprop} gets simplified in such a way that only the diagonal contributions $a_{\omega}\equiv a_{\omega,\omega}$ survive, while the term with $\mathcal{A}_{\omega,\omega'}$ drops out completely.
One should note that the periodicity $a_\omega = a_{\omega+n2\pi}, n\in \mathbb{N},$ is consistent with  quasienergies $\tilde\varepsilon_q$ being defined up to a shift in multiples of $2\pi$. Transitions caused by the coupling to the ancillas are thus well behaved in the Floquet sense.
While function $a_\omega$ captures the type of coupling to the ancilla qubits, positive real functions
\begin{align}
\new{g^s_{q,q'}= (1+\cos(q+q'))|\tilde{u}_{q'} \tilde{u}_q - \tilde{v}^{*}_{q'}\tilde{v}_q|^2,\quad 
g^{ca}_{q,q'}=(1+\cos(q'-q))|\tilde{u}_{q'} \tilde{v}_q - \tilde{v}_{q'} \tilde{u}_q|^2,}
\label{eq::formfunising}
\end{align}
take into account the transverse field Ising parameters. 

\begin{figure}[t!]
\centerline{\includegraphics[width=1\columnwidth]{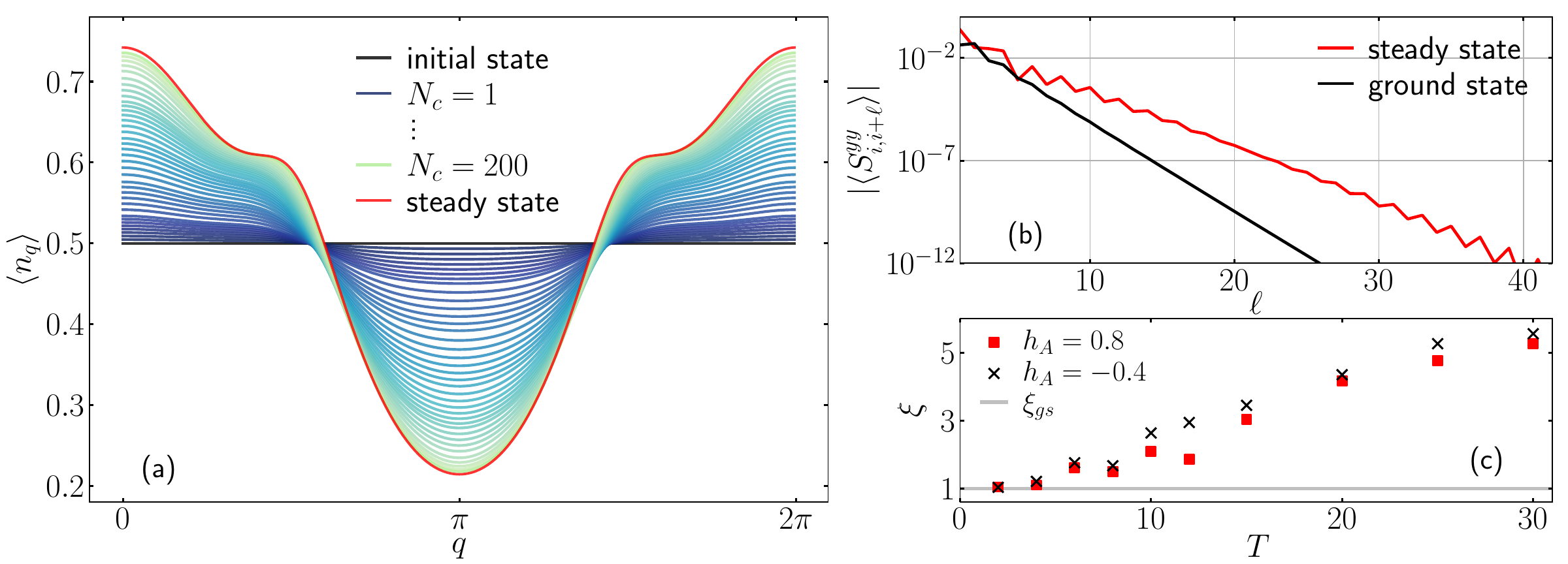}}
\caption{\new{(a)} Time evolution of the mode occupation from an initial infinite temperature state. A highly non-thermal steady state distribution is reached, which could be stabilized by the system-ancilla coupling in a digital quantum computer. Parameters: $J = 0.8, h=0.45, h_A = 0.8, T = 6$, $L=500, \lambda_{\tau} = \sqrt{\epsilon} = 0.1$. (b) Decay of correlations $|\ave{S^{yy}_{i,i+\ell}}|$, Eq.~\eqref{eq::isingC}, as a function of $\ell$ in the steady-state GGE and the ground state for the same parameters as in panel (a). As a signature of the stabilized non-thermal GGE, operators that overlap with local conserved quantities of transverse field Ising models show a slower decay of spatial correlations compared to the ground state. (c)~Different choices of system-ancilla coupling parameters (field $h_A$ and cycle duration $T$) yield different correlation lengths $\xi$. Quite generically, longer cycles lead to slower decay of spatial correlations and thus more non-thermal states. Other paramters are the same as in panel (a) and (b): $J = 0.8, h=0.45, L=500$.}
\label{fig::SCnq}
\end{figure}

We consider a time evolution from an infinite temperature state with $\mu_q=0$, which would locally describe an initial state in the digital quantum computer prepared by applying a few layers of (translationally invariant) random two-site gates on some product state \cite{schmitt22}. In Fig.~\ref{fig::SCnq}, we show the (zeroth order) GGE evolution from this state for parameters $J = 0.8$, $h=0.45$, $h_A = 0.8$, $T = 6$, $L=500$ and constant $\lambda_\tau=\sqrt{\epsilon}=0.1$ for which $a_\omega=\epsilon\sin^2(\frac{T}{2}(\omega-\pi h_A))/\sin^2(\frac{1}{2}(\omega-\pi h_A))$.
We see that out of a featureless infinite temperature state, some non-thermal features quickly start to appear, and the steady state is reached after approximately $N_c\sim 100$ reset cycles for the above parameters. The steady state itself has a clearly non-thermal occupation of eigenmodes, which depends on the system parameters $J,h$ via functions \new{$g^s_{q,q'}$, $g^{ca}_{q,q'}$} and on the parameters of system-ancilla coupling $h_A, T$ via function $a_\omega$.
Our main observation is that without a careful tuning of the ancilla parameters and coupling operators $Q_j, A_j$, weak constant coupling $\lambda_\tau=\sqrt{\epsilon}\ll 1$ of the integrable evolution to the ancilla qubits stabilizes a highly non-thermal population of eigenmodes. For the purpose of dissipative cooling, one has to tune the protocol such that (single- or multi-) quasiparticle decay processes (for our case last term in Eq.~\eqref{eq::nqdotGoogle}) are enhanced, as done in Refs.~\cite{mi23, lloyd2024}

While mode occupation clearly exposes the non-thermal nature of the stabilized state, it cannot be measured directly in a digital quantum computer, which has access only to local observables in the spin language. Local observables, which can expose the non-thermal nature of the stabilized state, are observables that strongly overlap with the local conserved quantities of the transverse field Ising model in the spin language \cite{essler16, grady82},
\begin{align}\label{eq::isingC}
C_0&= H_0,\quad
C_2=\sum_j 
J S^{xx}_{_{j,j+2}}
- h S^{yy}_{j,j+1} 
- h S^{xx}_{j,j+1}
- J \sigma^z_{j} \\
 C_{2\ell>2}&=  \sum_j 
J  S^{xx}_{_{j,j+\ell+1}} 
- h_x  S^{yy}_{_{j,j+\ell}}
- h_x  S^{xx}_{_{j,j+\ell}}
+J  S^{yy}_{_{j,j+\ell-1}}, \quad
C_{2\ell-1}= J \sum_j  S^{yx}_{_{j,j+\ell}} -  S^{xy}_{_{j,j+\ell}}. \notag
\end{align}
where $S^{\alpha\beta}_{i,j}=\sigma^\alpha_i \sigma^z_{i+1} \dots\sigma^z_{j-1} \sigma^\beta_{j}$. Observables $S^{xx}_{i,j}$ and $S^{yy}_{i,j}$ are experimentally accesible and have been measured also in Ref.~\cite{mi23}. 
In Fig.~\ref{fig::SCnq}(b), we plot $|\ave{S^{yy}_{i,i+\ell}}|$ in the GGE steady state as a function of $\ell$ and compare it to expectation values in the ground state ($\ave{n_q}=0$). Because we choose a non-critical set of system parameters, $J=0.8, h=0.45$, ground state and steady state correlations are decaying exponentially. The smoking gun for the GGE stabilization is a slow decay of spatial correlations in the steady state, {$|\ave{S_{i,i+\ell}^{yy}}| \sim e^{-\ell/\xi}$, which is even slower than the ground state one, $\xi>\xi_{gs}$. For the chosen Ising parameters $J$ and $h$, $\xi_{gs} \approx 1$, which is not true generically. 
In Fig.~\ref{fig::SCnq}(c) we show that with different choices of system-ancilla coupling parameters, one can tune the correlation length $\xi$. Quite generically, a longer reset time $T$ induces a slower (more non-thermal) decay of spatial correlations. However, this requires a larger number of gates and in total a longer circuit, which comes with a stronger influence of the inherent noise. 


A slow decay of correlations in the steady-state for the operators that overlap with the conserved quantities of the transverse field \new{Ising model is a direct consequence of reviving effects of integrability. In this case, the integrability is perturbed but also revived due to the dissipative coupling to ancilla qubits. Same would hold in the presence of weak additional unitary integrability breaking or native noise, which would introduce to Eq.~\eqref{eq::nqdotGoogle}  additional scattering terms. These would change the steady state and the particle distribution quantitatively, but would preserve it's non-thermal, structured nature as long as the inherent noise is not too strong and does not bring the system toward a featureless infinite temperature state.} 

\new{Since all above calculations were performed within the GGE approximation, using the scattering formalism, we compare in App.~\ref{app::TnExact} the time evolution of $\ave{S_{i,i+1}^{xx}}$ correlators calculated from the tensor network representation of the exact state and the GGE Ansatz. Similar to the continuous-time setup, exact results approach the GGE expectation values as $\lambda_{\tau}=\sqrt{\epsilon}$ is reduced.}

\section{Conclusions}\label{sec::conclusions}
We derived an effective description of non-interacting integrable many-body systems, \new{relaxing to highly non-thermal steady states due to the interplay of integrability and weak coupling to baths. We} discussed how such setups could be realized with digital quantum computers, such as superconducting circuits \cite{mi23} or trapped ions \cite{ringbauer22}. 
\new{Numerically supported by a comparison to the exact dynamics performed with tensor network TEBD simulation, we claim} that generalized Gibbs ensembles with generalized chemical potentials associated with mode occupation operators offer \new{a zeroth-order description and} a compact interpretation of time evolution and stabilized steady states. Namely, weak integrability breaking perturbations cause scattering between Bogoliubov quasiparticles, and we derived a generalized scattering theory, reminiscent of the Boltzmann equations, which yields the time-dependent eigenmode population, see also \cite{bouchoule20,perfetto23,rowlands23,lehr25,riggio24,lloyd2024}. 
The non-thermal nature of the stabilized steady states can be inferred from the structured distribution over eigenmodes, which is related to the transition rates between different quasiparticles caused by the integrability-breaking bath coupling. 


We proposed how to use digital quantum computers to realize such highly non-thermal GGEs due to proximity to integrability. There, driven-dissipative effects can be implemented by weakly coupling the system and ancilla qubits and resetting the latter at the end of every reset-cycle \cite{mi23}. We derived the effective system's density matrix time evolution for such a Floquet-reset protocol. By optimizing the system-ancilla coupling strength, Refs.~\cite{mi23, lloyd2024} recently prepared correlated many-body states close to the ground state. 
Our example shows that integrable systems that are weakly but generically coupled to ancilla qubits are actually prone to relax to highly non-thermal and structured GGEs. We comment on how such a highly non-thermal nature could be detected by measuring the decay of correlations that are slower than in the ground state.
\new{If not dominant,} additional native noise of the proposed platform is not detrimental for the observation of desired physics; while it would alter the time evolution and the steady state momentum occupations, it would preserve its highly non-thermal nature.
A digital quantum computer realization of our proposal would support a series of theory works \cite{lange17,lenarcic18,lange18,reiter21,schmitt22,ulcakar23} revealing a peculiar nature of nearly integrable models that can show a strong non-linear \new{steady state} response to weak coupling to non-thermal baths \new{and} would demonstrate that a similar activation of integrable effects could be possible also in nearly integrable materials \cite{lange17, scheie21,scheie22}.

Note: During the preparation of this manuscript, a related work appeared on arXiv \cite{lloyd2024}, optimizing the cooling process and interpreting the dissipative steady state preparation of Ref.~\cite{mi23} in terms of the scattering theory equivalent to ours.

\section*{Acknowledgements}
We thank R. Sharipov, G. Lagnese, J. Lloyd,  A. Rosch,  T. Prosen and M. \v{Z}nidari\v{c} for useful discussions. 

\paragraph{Funding information}
We acknowledge the support by the J1-2463 and N1-0318 projects of the Slovenian Research Agency, the QuantERA grants QuSiED and T-NiSQ by MVZI and QuantERA II JTC 2021 (ZL); the P1-0044 program of the Slovenian Research Agency and ERC StG 2022 project DrumS, Grant Agreement 101077265 (ZL and IU).

\begin{appendix}
\numberwithin{equation}{section}

\section{Comparison of approaches for the steady state calculation}\label{sec::approaches}
Since different approaches to nearly integrable, weakly dissipative system are still rather new \cite{lange17,lenarcic18,lange18,ulcakar23,gerbino23,perfetto23,rowlands23,lehr25,bouchoule20,riggio24} and not necessarily fully optimal, we review them here and compare their complexity:\\
\noindent
\underline{(1) Direct steady state calculation:} If aiming directly for the steady state, one can find the steady state Lagrange parameters $\mu_q(t\to\infty)$ from the stationarity condition $\ave{\dot{n}_{q}}=0$, Eq.~\eqref{eq::nkdot}, for all momenta. If considering a system of $L$ sites with $L$ mode occupation operators, the complexity of such a root finding procedure is $\mathcal{O}(L^{b+1})$, where $\mathcal{O}(L)$ is the complexity of evaluating the expression $\ave{\dot{n}_{q}}$ and $\mathcal{O}(L^{b})$ is the complexity of finding the root for $L$ variables. For example, $b=2$ for Powell method \cite{press07}.\\
\underline{(2) Iterative steady state calculation:} 
In Ref.~\cite{ulcakar23}, we developed an iterative approach for constructing the conserved quantities $\tilde{C}_{k}$, which play the leading role in a truncated generalized Gibbs ensemble description of the steady state, 
\begin{equation}
\rho_{\boldsymbol{\tilde{\lambda}}}^{(k)} =  \frac{e^{-\sum_{k'=0}^k \tilde\lambda_{k'}^{(k)} \tilde{C}_{k'}}}{\tr\Big[e^{-\sum_{k'=0}^k \tilde\lambda_{k'}^{(k)} \tilde{C}_{k'}}\Big]}    
\end{equation}
As the zeroth approximation to the steady state a Gibbs ensemble is taken, $\rho_{\boldsymbol{\tilde{\lambda}}}^{(0)}\propto e^{-\tilde\lambda_{0}^{(0)} H_0}$, with the zeroth iterative conserved quantity being the Hamiltonian, $\tilde{C}_0=H_0$. In next iterative steps, the $k$th iterative conserved quantity is constructed in the basis $Q_m$, $[H_0,Q_m]=0$ as  
\begin{equation}\label{eq::iterA}
\tilde{C}_k = 
\mathcal{N}_k^{-1}
\sum_m w^{(k)}_m Q_m, \quad
w^{(k)}_m= - \sum_n \left(\chi_{(k-1)}^{-1}\right)_{mn} \tr[Q_n \cl{D} \rho_{\boldsymbol{\tilde\lambda}}^{(k-1)}]. 
\end{equation}
For the non-interacting $H_0$, a natural choice is $Q_m=n_m$, the basis of mode occupation operators. In this case, 
$(\chi_{(k)})_{m,n} = \ave{Q_m Q_{n}} - \ave{Q_m} \ave{ Q_{n}}= e^{-\mu_m^{(k)}}/(1+e^{-\mu_m^{(k)}})^2 \delta_{m, n}$ the susceptibility matrix is diagonal, which further reduces the complexity of performing the iterative procedure. Here, $\mu_m^{(k)}$ is an effective Lagrange parameter associated to the mode occupation operator $n_m$ at $k$th iterative step, $\mu_m^{(k)} = \tilde\lambda_{0}^{(k)} \varepsilon_m + \sum_{k'=1}^k \mathcal{N}_{k'}^{-1} \tilde\lambda_{k'}^{(k)} w^{(k')}_m$, and $\varepsilon_m$ is the dispersion.
The approximation to the steady state is established by finding  
$\{\tilde\lambda_{k'}^{(k)}\}$
for 
$\rho_{\boldsymbol{\tilde{\lambda}}}^{(k)}\propto e^{-\sum_{k'=0}^k \tilde\lambda_{k'}^{(k)} \tilde{C}_{k'}}$
from the set of $k+1$ conditions $\ave{\dot{\tilde{C}}_{k'}}=0$, Eq.\eqref{eq::nkdot}, for $\{\tilde{C}_{k'}\}_{k'=0}^k$.
We set normalization $\mathcal{N}_k$ to be $1$, thereby absorbing it into the corresponding Lagrange parameters.

The complexity of the procedure scales as $\mathcal{O}(k^3 L)$ for the Powell method. If $k \sim \mathcal{O}(1)$ and small, for thermodynamically large systems, the iterative method is clearly advantageous to the previous approach.\\
\underline{(3) Truncated GGE (most local conserved quantities):} 
In principle, another possibility is the truncation in the Fourier modes of $\ave{n_q}$ or in the number of local conserved quantities $C_i$ of the spin model that are considered \cite{lange17,lange18,reiter21,pozsgay13,fagotti13gge2,essler16,pozsgay17}. $C_i$ are for the transverse field Ising model linearly related to the mode occupation operators as 
$C_{2\ell} = \sum_{q} \cos{(q\ell)} \varepsilon_{q} n_q$ for even ones ($C_0=H_0$) and as $C_{2\ell-1} = 2J\sum_{q} \sin{(q\ell)} n_q$ for odd ones \cite{fagotti13gge2}. If one includes only $N_i$ most local ones, $2\ell < N_i$, then the complexity of finding the truncated steady state GGE scales as $\mathcal{O}(L N_i^2)$.\\
\underline{(4) Time propagation:} As done in the main text, one can calculate the whole time evolution from some initial $\mu_q(0)$, using a discretized version of Eq.~\eqref{eq::mudot} and, for example, the Euler method. The complexity of such a calculation is $\mathcal{O}(N_t L)$, where $N_t$ is the number of steps needed to reach the steady state. If we aim to calculate the steady state, the initial $\mu_q(0)$ can be a guess for the steady state. On the other hand, if we aim to describe a realistic time evolution from a state $\ket{\psi_0}$, the initial $\mu_q(0)$ are given by the initial state through the condition 
$\ave{\psi_0|n_q|\psi_0}=\tr\left[n_{q} \frac{e^{-\sum_{q'} \mu_{q'}(0) n_{q'}}}{\tr[e^{-\sum_{q'} \mu_{q'}(0) n_{q'}}]}\right]$. However, this itself is a root-finding procedure which requires $\mathcal{O}(L^{b+1})$ steps.

\begin{figure}[t!]
\centerline{\includegraphics[width=1\columnwidth]{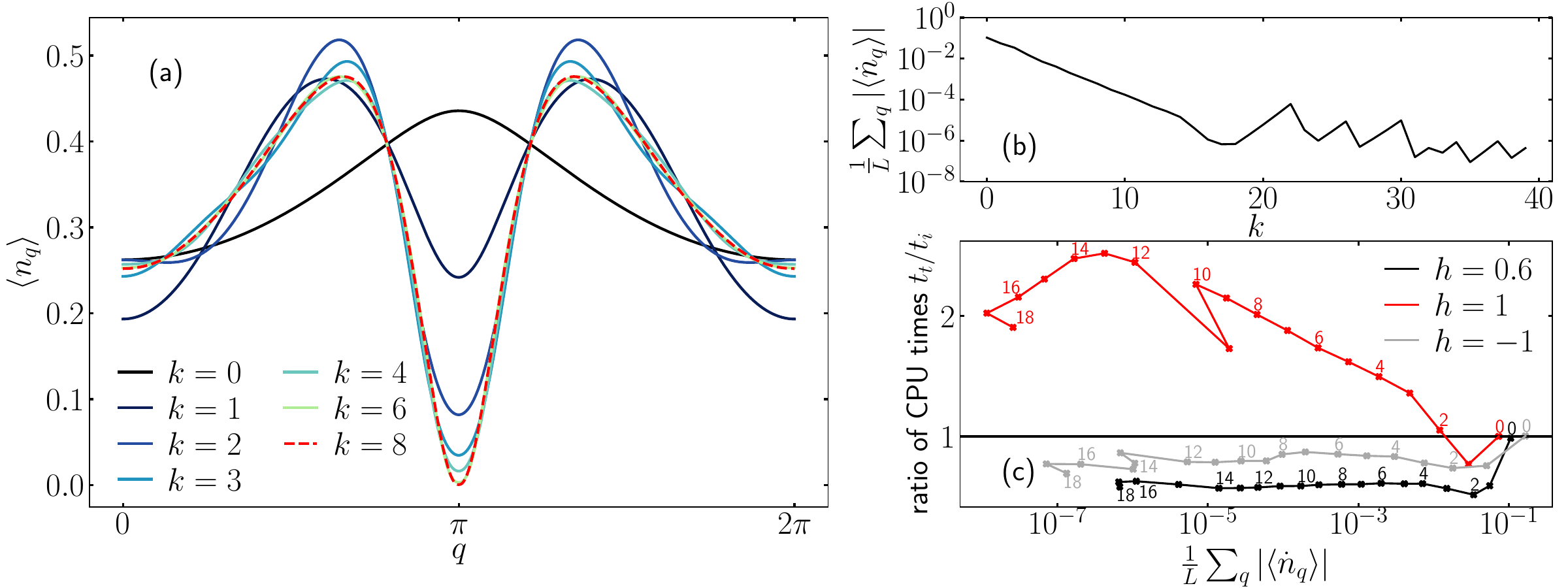}}
\caption{(a) Convergence to the steady state mode occupation at different iterative steps $k$. In the $k=0$ step, the steady state is approximated by a thermal state. In the following iterative steps, additional leading conserved operators are added to a truncated GGE. A decent convergence is obtained in finite number of steps. 
(b) After the initial improvement of results with increasing number of iterative steps, for chosen parameters, $k>18$ iterative steps fail to improve the results further. However, this happens in the regime where results are converged for all practical purposes.
Parameters: $J=1, h=0.6, L=10^5$.
(c) Ratio of computing times $t_t/t_i$, where $t_t$ corresponds to time evolution with $\epsilon\delta t =0.6$ and $t_i$ to calculation with the iterative scheme, as a function of $(1/L)\sum_q |\ave{\dot{n}_q}|$, characterizing the accuracy of steady state calculation. Points are labeled by the number of iterative step taken for $t_i$ calculation. The two methods are comparable. Which one is more efficient in absolute terms depends on parameters.  Parameters: $J=1, L=10^5$.}
\label{fig2}
\end{figure}

The approach (1) is clearly disadvantageous to others and will not be considered. Below we compare approach (4) to the iterative approach (2) from Ref.~\cite{ulcakar23}. We perform the comparison for the model introduced in Sec.~\ref{sec::model}, where the time-dependent calculation \eqref{eq::nkdot} has already been performed.

Fig.~\ref{fig2} shows results for the iterative steady state calculation, Eq.~\eqref{eq::iterA}. We start with an initial approximation in the form of a Gibbs ensemble, with Hamiltonian being the only conserved quantities. Then, we perform our iterative procedure for constructing a truncated GGE steady state description. 
The leading conserved quantities $\tilde{C}_k$, Eq.~\eqref{eq::iterA}, are a linear superposition of the basis mode occupation operators $n_q$ with weights selected by the dissipator.
Fig.~\ref{fig2}(a) shows momentum distributions obtained after $k$ iterative steps. The initial $k=0$ distribution corresponds to the thermal ensemble at a temperature that best represents the steady state, as obtained from a steady state rate equation for the energy. We observe that convergence to the steady state is obtained in a finite number of $k=8$ steps when we cannot discern this distribution from the ones of the following iterative steps.
In Fig.~\ref{fig2}(b), we push the number of iterative steps further, even though this is not needed for practical purposes. We observe that improvement is obtained only up to $k=18$ iterative steps. The reason might be that with further steps, we are not adding new direction to the GGE manifold or that we are dealing with extremely small weights in \eqref{eq::iterA} that can be numerically unstable and prone to errors. However, this problematic behavior appears in, for all practical purposes, an irrelevant regime.

In Fig.~\ref{fig2}(c) we compare the efficiency of the direct time propagation, Eq.~\eqref{eq::nkdot}, and the iterative approach, Eq.~\eqref{eq::iterA} by plotting the ratio of CPU times for the former vs the latter. We show that as a function of the average remaining flow of the mode occupations, $(1/L)\sum_q |\ave{\dot{n}_q}|$, characterizing how far from the steady state is the approximate description at a given iterative or finite time step. Fig.~\ref{fig2}(c) reveals that the two methods are comparable, as anticipated from the scaling arguments. Namely, the numerical complexity of time propagation scales as $\mathcal{O}(N_t L)$, where $N_t$ is the number of propagation steps, while the iterative method scales as $\mathcal{O}(k^3 L)$, where $k$ is the number of needed iterative steps. For the case studied, direct propagation can be performed at rather large $\epsilon \delta t=0.6$ time steps, meaning that the direct propagation is rather efficient. We could have gained some efficiency for the iterative method by not converging the steady state equations at intermediate iterative steps, however, we did not play with that knob. Which approach is quantitatively advantageous depends on the choice of parameters $J,h$.

In Fig.~\ref{fig1}(c) of the main text, we plot the steady state expectation values of local conserved quantitites $\ave{C_i}$, Eq.~\eqref{eq::isingC}. Since the steady state mode occupation is symmetric, $\ave{n_q} = \ave{n_{-q}}$, only parity-even conserved quantities have finite expectation values. Fig.~\ref{fig1}(c) reveals exponentially decaying contribution with growing support, which indicates that also more standard truncation using the most local conserved quantities is meaningful. If $N_i$ conserved quantities are used, the complexity of calculating the steady state scales as $\mathcal{O}(L N_i^2)$. Because we expect that our iterative construction is more efficient, we do not perform a detailed comparison.


Our main conclusion from this analysis is that a direct, steady state calculation of all Lagrange parameters for all the mode occupation operators from the stationarity condition of Eq.~\eqref{eq::nkdot} is the most costly ($\mathcal{O}(L^3)$) and should be avoided. Other approaches are comparable; which one is the most efficient depends on the model parameters. 

\section{Floquet transverse field Ising model}
\label{app::FTFI}
In this section, we discuss the generalized Bogoliubov rotation for the Floquet transverse field Ising model, 
\begin{equation}\label{eqS::US}
U_S = e^{-i\frac{\pi J}{2}\sum_{j}\sigma^x_j\sigma^x_{j+1}} e^{-i\frac{\pi h}{2}\sum_{j}\sigma^z_j}
= e^{-i H_{\text{FTFI}}},
\end{equation}
relevant for a digital quantum computer realization, Sec.~\ref{sec::SCQubits}. 
Using the Jordan-Wigner transformation, Eq.~\eqref{eq::JW}, the Fourier transform, Eq.~\eqref{eq::FTfermion}, and periodic boundary conditions, system's time evolution factorizes over momenta as 
\begin{equation}\label{eq::Usq}
U_S = \prod_{q \geq 0} e^{-i \Phi_q^\dagger X_q \Phi_q} e^{-i \Phi^\dagger_q Z_q \Phi_q},
\end{equation} 
with $\Phi_q = \{c_q, c_{-q}^\dagger \}^T$ representing the bispinor of fermionic operators in momentum space, Eq.~\eqref{eq::FTfermion} and $2 \times 2$ matrices 
\begin{equation}
X_q= \pi J
\begin{bmatrix}
\cos(q) & -\sin(q)\\
-\sin(q) & -\cos(q) 
\end{bmatrix}, \quad
Z_q = \pi h 
\begin{bmatrix}
1 & 0\\
0 & -1 
\end{bmatrix}.
\label{eq::matrices}
\end{equation}
Factorization \eqref{eq::Usq} is possible since $X_q$ commute amongst each other for positive momenta but not necessarily with their negative momenta counterparts.
Dispersion relation $\tilde\varepsilon_q$ and the Bogoliubov transformation are obtained by diagonalizing each $q$-block $e^{-i X_q} e^{-i Z_q }$ separately,
\begin{equation}
P^{-1} e^{-i X_q} e^{-i Z_q } P = \text{diag}[e^{-i \tilde \varepsilon_q},e^{i \tilde \varepsilon_{q}}], 
\label{eq::matrixdiag}
\end{equation}
yielding
\begin{equation}
\cos(\tilde\varepsilon_q) = \cos(\pi J) \cos(\pi h) - \sin(\pi J) \sin(\pi h) \cos(q).
\end{equation}
The Bogoliubov transformation, $\Phi_q^\dagger P = (d_q^\dagger, d_{-q})$, then takes a similar form as in the continuous-time propagation 
\begin{align}
    c_q &= \tilde{u}_q d_q - \tilde{v}^*_qd_{-q}^{\dagger},\
    \tilde u_q = \frac{\xi_q+ \tilde a_q}{\sqrt{2\xi_q(\xi_q + \tilde a_q)}}, \
    \tilde v_q = \frac{\tilde b_q}{\sqrt{2\xi_q(\xi_q + \tilde a_q)}}, \
    \xi_q = \sqrt{\tilde a_q^2 + |\tilde b_q}|^2\\
    \tilde a_q &= \sin(\pi J) \cos(\pi h) \cos(q) + \cos(\pi J) \sin(\pi h), \
    \tilde b_q = -e^{-i \pi h} \sin(\pi J) \sin(q).\notag
\end{align}
The system's unitary time propagator in the diagonal form then equals
\begin{equation}
    U_S = e^{-i\sum_q \tilde{\varepsilon}_q (d_q^{\dagger}d_q - \frac{1}{2})}.
\end{equation}
Above we were able to consider the diagonalization of one $q$-block $e^{-i X_q} e^{-i Z_q} = e^{-i H_{q, \mathrm{FTFI}}}$ from Eq.~\eqref{eq::matrixdiag} as a matrix and not as an operator, $e^{-i \Phi_q^\dagger X_q \Phi_q} e^{-i \Phi^\dagger_q Z_q \Phi_q} = e^{-i \hat{H}_{q, \mathrm{FTFI}}}$, since we can show that the Floquet Hamiltonian is of form $\hat{H}_{q, \mathrm{FTFI}} = \Phi_q^\dagger H_{q, \mathrm{FTFI}} \Phi_q$. This is shown by realizing that for any matrices $\Phi_q^\dagger A \Phi_q$ and $\Phi_q^\dagger B \Phi_q$, where $\Phi_q = \{c_q, c_{-q}^\dagger \}^T$ is the fermionic bispinor in momentum space, the following commutation relation holds: $[\Phi_q^\dagger A \Phi_q , \Phi_q^\dagger B \Phi_q] = \Phi_q^\dagger [A, B] \Phi_q$. From this, it follows that finding the effective Floquet transverse field Ising Hamiltonian for momentum $q$ in the operator form is equivalent to finding it in the matrix form (e.g., via the Baker-Hausdorff-Campbell formula) and applying bispinor operator $\Phi_q^\dagger$ left and $\Phi_q$ right of the Floquet Hamiltonian matrix.

\section{Lindblad evolution of system's density matrix in a digital quantum computer propagation}
\label{app::rhoTEvo}
Here, we derive the discrete time evolution of the system's density matrix, Eq.~\eqref{eq::Tprop}, for the Trotterized gate propagation in a digital quantum computer, where dissipation is due to the coupling and reset of ancillary qubits.
We first derive the equation of motion for a general case and then narrow it down for our model. 

The system's Trotterized time evolution is for one step given by a unitary $U_S$. Simultaneously, Trotterized time evolution on ancilla qubits is performed by $U_A$.
This is always followed by a weak hermitian system-ancilla coupling 
\begin{align}
\label{eq::interactiondef}
U_{SA, \tau} &= \prod_j e^{ -i \lambda_{\tau} Q_j \otimes A_j}
\approx e^{ -i \lambda_{\tau}\sum_j Q_j \otimes A_j - \frac{1}{2}\lambda^2_{\tau}\sum_{j, j'} [Q_j, Q_{j'}] \otimes A_jA_{j'}} 
\equiv e^{ -i W_{\tau}},
\end{align}
where $Q_j$ and $A_j$ are hermitian operators acting on system and ancilla qubits respectively. We assumed that $A_j$ are single site operators, while $Q_j$ can be multi-site operators. We have introduced an effective coupling Hamiltonian $W_{\tau}$ at time step $\tau\le T$, which includes the first and second order terms of the expansion in $\lambda_{\tau}\ll 1$. Higher order terms are neglected.
One cycle contains $T$ system-ancilla-coupling propagations
\begin{equation}
U_T=U_{SA, T} U_A U_S \cdots U_{SA, 1} U_A U_S, 
\label{eq::Stimeprop}
\end{equation}
\new{followed by a reset of ancilla qubits to a chosen spin state.}

Following Ref.~\cite{lidar2019}, we derive the system's density matrix time evolution in the interaction picture, which is slightly non-standard due to the Trotterized nature of the setup. 
The interaction picture propagator for one cycle (before the reset) equals
\begin{equation} 
\mathcal{U}_T \equiv U_0^{-T}U_T = \hat{\mathcal{T}} e^{-i \sum_{\tau=1}^{T} W_{I \tau}}, \, 
U_0=U_A U_S,
\label{eq::intpicprop}
\end{equation}
where $W_{I\tau} = U_0^{-\tau} W_{\tau} \, U_0^{\tau}$ is the first and second order of the effective coupling Hamiltonian \eqref{eq::interactiondef} propagated in the interaction picture for $\tau$ steps and $\hat{\mathcal{T}}$ is the time ordering operator. In App.~\ref{app::timeprop}, we prove Eq.~\eqref{eq::intpicprop}. 

\new{Due to the reset}, the whole density matrix operator has a product form at the end of each cycle,
\begin{equation}
\rho_{I}(N_c) = \rho_{S,I}(N_c)\otimes \prod_j \tilde{P}_j,
\end{equation}
\new{where $\tilde{P}_j$ is the projection of $j$th ancilla qubit on the chosen state.}
One reset cycle evolution of the system's density matrix $\rho_{S,I}$, obtained by tracing out the ancilla qubits, is approximated to second order in coupling strength $\lambda_{\tau}$ by
\begin{align}
\rho_{S,I}(N_c+1) - \rho_{S,I}(N_c)
&=\tr_A \Big(\mathcal{U}_T\rho_{I}(N_c) \, \mathcal{U}_T^{\dagger}\Big) - \rho_{S,I}(N_c)\notag\\
&\approx 
-i\tr_A\Big(\sum_{\tau=1}^{T}[W_{I\tau},\rho]\Big)
-\tr_A\Big( 
\sum_{\tau=1}^{T} \sum_{\tau'=1}^{\tau} [\lambda_\tau V_{I\tau},[\lambda_{\tau'} V_{I\tau'}, \rho_I(N_c)]] \Big) \label{eq::rholin}\\
&= \sum_{\tau=1}^{T} \sum_{\tau'=1}^{\tau} \sum_{i,j} \lambda_\tau \lambda_{\tau'}
\Big( \big(Q_{i,I\tau} Q_{j,I\tau'}\rho_{S,I}(N_c)-Q_{j,I\tau'} \rho_{S,I}(N_c) Q_{i,I\tau}\big) \mathcal{A}_{i, j, \tau,\tau'} \notag \\
&\hspace{3cm} +\big( \rho_{S,I}(N_c) Q_{j,I\tau'} Q_{i,I\tau}-Q_{i,I\tau} \rho_{S,I}(N_c) Q_{j,I\tau'}\big) \mathcal{A}^*_{i, j, \tau,\tau'}
\Big). \notag
\end{align}
The linear term in \eqref{eq::rholin} can be set to zero by shifting the $A_j$ operators \cite{lidar2019}. But for our choice $A_j=\tilde{\sigma}_j^x$ it vanishes trivially since $\tr_A\big(U_0^{-\tau} A_j U_0^\tau (\tilde{\mathbb{1}} - \tilde{\sigma}_j^z)\big)=0$.
In a compact notation, all effects of ancilla qubits (unitary evolution $U_A$, the coupling operator acting on ancilla $A_j$ and the resetting of ancillas), is represented by
\begin{equation}
    \mathcal{A}_{\tau,\tau', i, j} = \tr_A \Big[\prod_k \tilde{P}_k \, A_{i,I\tau} A_{j,I\tau'} \Big].
\end{equation}

While the above derivation and expressions are generic, we now simplify them further by turning to our model with ancilla propagator $U_A= e^{-i\frac{\pi h_A}{2}\sum_j \tilde{\sigma}_j^z}$, ancilla term $A_j = \tilde\sigma^x_j$ and resetting \new{protocol to all down state} $\tilde{P}_k=\frac{1}{2}\big(\tilde{\mathbb{1}} - \tilde{\sigma}_k^z\big)$,
\begin{align}
    \mathcal{A}_{\tau,\tau', i, j}
    = \tr_A \Big[\prod_k \frac{1}{2}\big(\tilde{\mathbb{1}} - \tilde{\sigma}_k^z\big) A_{i,I\tau} A_{j,I\tau'} \Big] \delta_{i,j}
    =e^{2ih_A(-\tau + \tau')}\delta_{i,j} 
    \equiv\mathcal{A}_{\tau,\tau'}. 
\end{align}
It is more convenient to represent the coupling operator $Q_j$ in terms of transitions it causes. Therefore we introduce
\begin{equation}
Q_{j,\omega} =  \sum_{\alpha, \beta,\\ \tilde{E}_\beta - \tilde{E}_\alpha = \omega} \ket{\alpha}\bra{\alpha} Q_j \ket{\beta}\bra{\beta},    
\end{equation}
which represents $Q_{j}$ operator projected between many-body eigenstates of $H_{\text{FTFI}}$ that differ in energy for $\omega$. Here, $\ket{\alpha}$ is a many-body eigenstate of the systems's unitary $U_S$ with a corresponding eigenvalue $e^{-i\tilde{E}_{\alpha}}$, where $\tilde{E}_{\alpha}$ is called the quasi-energy of the Floquet Hamiltonian $H_\text{FTFI}$. Then
\begin{equation}
 Q_{j,I\tau} 
    = \sum_\omega U_0^{-\tau} Q_{j,\omega} U_0^{\tau},  
 = \sum_\omega e^{-i\omega \tau } Q_{j,\omega},   
\end{equation}
Accompanying, we introduce the ancilla correlations functions represented in the frequency space
\begin{align}
\mathcal{A}_{\omega,\omega'}
&=\sum_{\tau=1}^{T} \sum_{\tau'=1}^{\tau} \lambda_\tau \lambda_{\tau'} e^{i\omega' \tau - i\omega \tau'}  \mathcal{A}_{\tau,\tau'}
= a_{\omega,\omega'} - \mathcal{A}^*_{\omega',\omega} \notag \\
a_{\omega,\omega'}
&=\sum_{\tau=1}^{T} \lambda_\tau e^{i(\omega'-\pi h_A)\tau} \sum_{\tau'=1}^{T} \lambda_{\tau'} e^{-i(\omega-\pi h_A)\tau'}
\end{align}
Putting all these together, we derive a compact form 
\begin{align}
&\rho_{S,I}(N_c+1) - \rho_{S,I}(N_c)=
\sum_{j,\omega, \omega'} 
-i\, {\rm Im}(\mathcal{A}_{\omega,\omega'}) [Q_{j,\omega'}^\dagger Q_{j,\omega}, \rho_{S,I}(N_c)] \notag \\
&\hspace{5cm} + a_{\omega,\omega'}\Big(Q_{j,\omega} \rho_{S,I}(N_c) Q_{j,\omega'}^\dagger - \frac{1}{2} \{Q_{j,\omega'}^\dagger Q_{j,\omega}, \rho_{S,I}(N_c)\}\Big).
\end{align}
To obtain the propagation equation  \eqref{eq::Tprop} presented in the main text, we approximate the system's density matrix with a GGE ansatz that, notably, does not evolve under $U_0$, making the transformation back to the Schrödinger picture trivial.

\section{Floquet interaction picture time propagator}
\label{app::timeprop}
In this section, we show that $\mathcal{U}_T$, Eq.~\eqref{eq::intpicprop}, really is the interaction picture propagator for one cycle consisting of $T$ system-ancilla-coupling propagations.

The Schr{\"o}dinger picture propagator \eqref{eq::Stimeprop} can be written as
\begin{align}
    U_T&=U_{SA, T} U_A U_S \cdots U_{SA, 1} U_A U_S, \\
    &=U_{SA, T} U_0 U_{SA, T-1} U_0\cdots U_{SA, 1} U_0 \notag\\
    &= U_0^T U_0^{-T} U_{SA, T} U_0 U_0^{T-1} U_0^{-(T-1)} U_{SA, T-1} U_0\cdots U_0 U_0^{-1} U_{SA, 1} U_0 \notag \\
    &= U_0^{T} U_{I, SA, T} U_{I, SA, T-1} \dots U_{I, SA, 1} \notag
\end{align}
Above we introduce the interaction picture coupling propagator $U_{I, SA, \tau} =  U_0^{-\tau} U_{SA, \tau} U_0^{\tau} \approx e^{-i W_{I \tau}}$ at step $\tau$. The interaction picture time propagator for one reset cycle of length $T$ is then 
\begin{equation}
    \mathcal{U}_T = U_0^{-T}U_T = e^{-i W_{I T}} e^{-i W_{I T-1}} \dots e^{-i W_{I 1}}= \hat{\mathcal{T}} e^{-i \sum_{\tau = 1}^T W_{I \tau}}
\end{equation}
In the last step, we used the following property of the time ordering operator: $e^{-i \hat{O}(t_2)}e^{-i \hat{O}(t_1)} = \hat{\mathcal{T}} e^{-i (\hat{O}(t_2)+\hat{O}(t_1))}$ for any operator $\hat{O}(t)$, if $t_2 > t_1$. Thus, we have shown that Eq.~\eqref{eq::intpicprop} holds.

\section{Examples and symmetries in the Trotterized setup}
In order to illustrate the variety of different non-thermal steady states stabilized, we show here a few examples of steady state mode occupations that were considered to demonstrate anomalously long spatial correlations in the main text, Fig.~\ref{fig::SCnq}(c). In Fig.~\ref{fig::SCnqApp}(a), we show the full time evolution of the mode occupation from the initial infinite temperature state, for $J = 0.8, h = 0.45, h_A = -0.4, T = 6$. It is interesting to observe that even though these parameters yield a comparable correlation length $\xi$ for the decay of spatial correlations in Fig.~\ref{fig::SCnq}(c) as $h_A = 0.8$, the steady state distribution is completely different from the distribution at $h_A=0.8$ shown in the main text, Fig.~\ref{fig::SCnq}.  

\begin{figure}[t!]
\includegraphics[width=1\columnwidth]{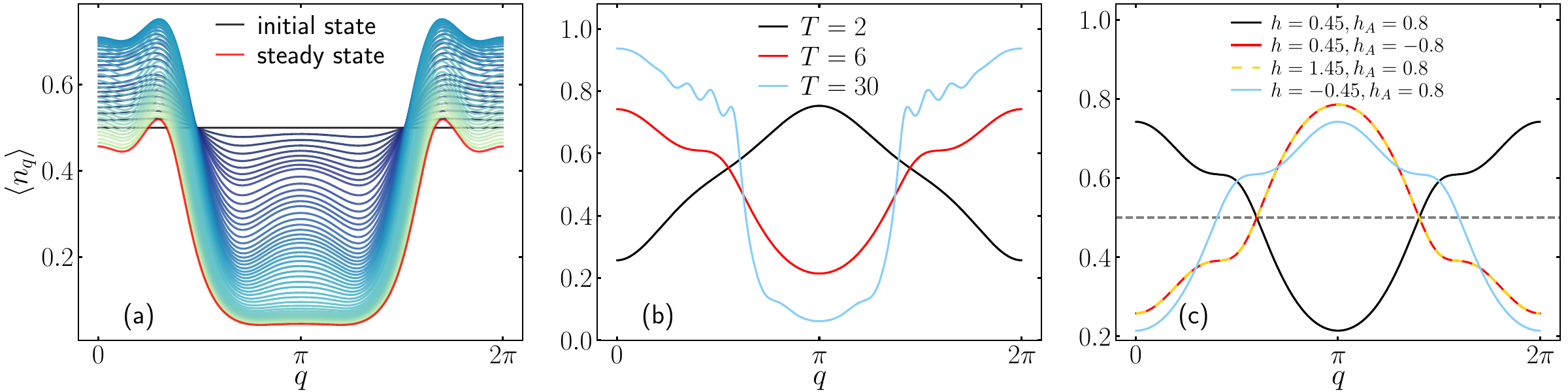}
\caption{
(a) Time evolution of the mode occupation from an initial infinite temperature state. Evolution correspond to the system-ancilla coupling in a digital quantum computer at parameters: $J = 0.8$, $h=0.45$, $h_A = -0.4$, $T = 6$, $L=500$, $\lambda_{\tau} = \sqrt{\epsilon} = 0.1$.
(b) Steady state distributions of the mode occupation for different lengths of the reset cycle $T$. Parameters: $J = 0.8$, $h=0.45$, $h_A = 0.8$, $L=500$, $\lambda_{\tau} = \sqrt{\epsilon} = 0.1$ and $T = 2, 6, 30$.
\new{(c) Steady state mode occupations under different symmetry transformations of the model. Taking $h_A \rightarrow -h_A$ or $h \rightarrow h+1$ will invert the steady state population, whereas $h \rightarrow -h$ will shift momentum by $\pi$. Parameters: $J = 0.8, T = 6, L=500$, $\lambda_{\tau} = \sqrt{\epsilon} = 0.1$.}
}
\label{fig::SCnqApp}
\end{figure}

In Fig.~\ref{fig::SCnqApp}(b), we show the steady state distributions of momentum occupations at three different lengths of the reset cycle, $T=2,6,30$, again for parameters $J = 0.8$, $h=0.45$, $h_A = 0.8$, $L=500$, $\lambda_{\tau} = \sqrt{\epsilon} = 0.1$ shown in the main text in Fig.~\ref{fig::SCnq}(c). Consistently with results from the main text, longer reset-cycles lead to more clearly non-thermal steady states yielding longer spatial correlations in $|\ave{S_{i,i+\ell}^{yy}}|$.


The equation of motion for the mode occupation,
\new{\begin{align}\label{eq::nqAPP}
&\ave{n_q(N_c+1)}-\ave{n_q(N_c)}
= \frac{2}{L} \sum_{q'} 
g^s_{q,q'} \big(\langle 1 - n_q \rangle \langle n_{q'} \rangle a_{\varepsilon_{q'}-\varepsilon_q} 
- \langle n_q \rangle \langle 1 - n_{q'} \rangle a_{\varepsilon_q-\varepsilon_{q'}}\big)   \\
&\hspace{5.3cm}+g^{ca}_{q,q'}\big(\langle 1 - n_{q} \rangle \langle 1 - n_{q'} \rangle a_{-\varepsilon_{q'}-\varepsilon_q} \notag - \langle n_q \rangle \langle n_{q'} \rangle a_{\varepsilon_{q'}+\varepsilon_q}\big), \notag
\end{align}}
has certain symmetries, which imply symmetric relations also for the steady state occupations.
\new{First, the equations of motion are invariant under shifting Ising parameters $J, h$ and the bath field $h_A$ by multiples of $2$.
Flipping the sign of the bath field results in the steady state occupations being reflected across the infinite temperature value $\ave{n_q}_{SS}(-h_A)= 1 - \ave{n_q}_{SS}(h_A)$. This is due to the exchange of roles of $a_{\omega}$ in the first and second term, as well as in the third and fourth term in Eq.~\eqref{eq::nqAPP}, since $a_{\omega}(-h_A)=a_{-\omega}(h_A)$. This results in $\ave{S_{i,i+\ell}^{yy}}_{SS}(-h_A)=-\ave{S_{i,i+\ell}^{yy}}_{SS}(h_A)$.
Likewise, if we shift one of the Ising parameters $J$ or $h$ by an odd integer, the steady state occupations become reflected across the infinite temperature value: $\ave{n_q}_{SS}(J+1)= 1 - \ave{n_q}_{SS}(J)$ and $\ave{n_q}_{SS}(h+1)= 1 - \ave{n_q}_{SS}(h)$. This occurs for the same reason as before; the roles of $a_{\omega}$ are exchanged, whereas $g^s_{q,q'} \text{and } g^ac_{q,q'}$ remain unchanged.
Lastly, if we flip the sign of one of the parameters $J$ or $h$, the steady state occupations shift by $\pi$: $\ave{n_q}(-h)=\ave{n_{q+\pi}}(h)$ and $\ave{n_q}(-J)=\ave{n_{q+\pi}}(J)$. In Fig.~\ref{fig::SCnqApp}(c) we show some of the discussed symmetries.
}

\section{Comparison of GGE ansatz and exact calculation for the Trotterized setup}
\label{app::TnExact}

\begin{figure}[t!]
\includegraphics[width=1\columnwidth]{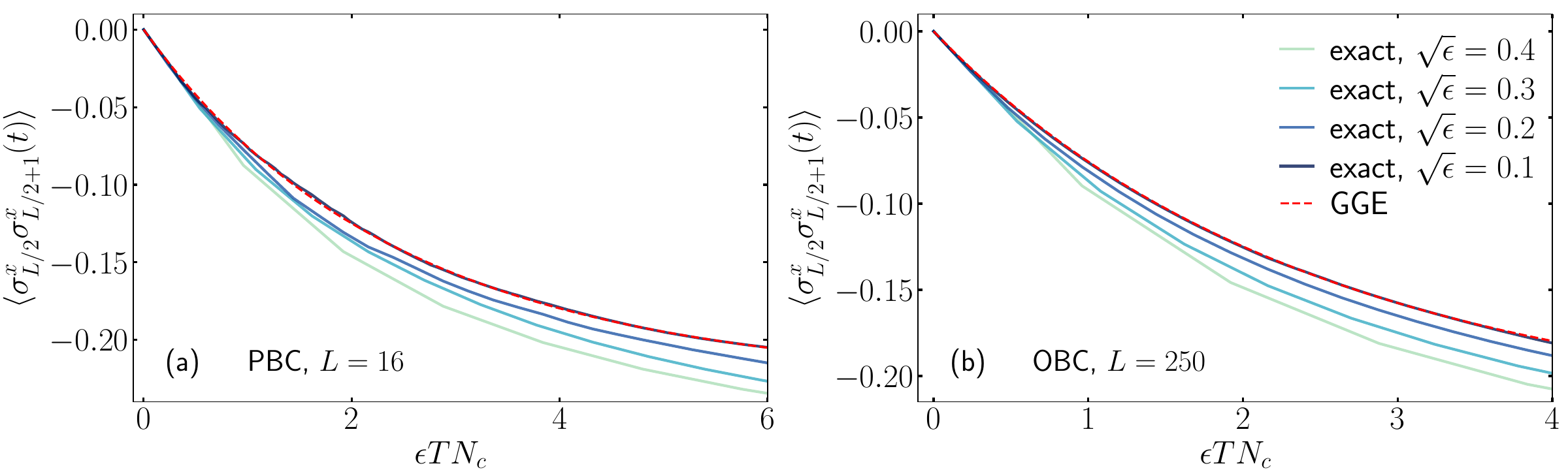}
\caption{\new{Comparison of exact and GGE Ansatz dynamics for observable $\ave{\sigma^x_{L/2} \sigma^x_{L/2+1}}$ as a function of cycle number $N_c$ scaled with square of system-ancilla coupling $\lambda_\tau^2=\epsilon$ and the reset period $T$ on (a)~$L=16$ system qubits in a circuit with periodic boundary conditions and (b) $L=250$ with open boundary conditions. Exact evolution is obtained with tensor network representation of the circuit from Fig.~\ref{figSC} at four different coupling strengths $\lambda_{\tau} = \sqrt{\epsilon } =  0.1, 0.2, 0.3, 0.4$.
Other parameters: $J = 0.8, T = 4$, $600\le \chi\le 800$ for PBC case and $\chi=300$ for OBC case.}}
\label{fig::TNExactApp}
\end{figure}

\new{In Fig.~\ref{fig::TNExactApp}, we compare dynamics of expectation values $\ave{\sigma^x_{L/2} \sigma^x_{L/2+1}}$ calculated with a tensor network representation of the dissipative circuit from the main text,  Fig.~\ref{figSC}, and via the zeroth order approximation to the dynamics given by the GGE Ansatz. We initialize the system in the infinite temperature state $\rho_S(0)=\mathbb{1}$ and ancilla qubits in the spin down state.
In Fig.~\ref{fig::TNExactApp}(a) we plot results for the circuits with periodic boundary conditions (PBC) used in the main text with $L=16$ system qubits and different coupling strengths $\lambda_{\tau} = \sqrt{\epsilon } = 0.1, 0.2, 0.3, 0.4$. By decreasing the coupling strength, results converge towards the GGE solution, confirming its validity also for the Trotterized setup. In order to achieve bond-dimension converged results for the shown cycle numbers at PBC, rather large bond dimensions have to be used  $600\le \chi\le 800$. In Fig.~\ref{fig::TNExactApp}(b), we perform the comparison of PBC GGE result and the open boundary conditions (OBC) tensor network results, where we can reach larger number of system qubits $L=250$ at manageable bond dimension $\chi=300$. For OBC circuits, boundary effect can be quite prominent, so one needs larger system sizes for convergence towards the GGE result. Since the reset protocol does not induce currents, expectation values of local operators in the bulk are not expected to be influenced by the choice of boundary conditions on large enough systems and can be compared to the GGE results at PBC. To summarize, both examples show that GGEs give a good approximation to the exact dynamics at small enough coupling between the system and ancilla qubits. To achieve convergence at even deeper circuits and $\lambda_\tau=0.1$, larger bond dimensions would be needed. Since the presented results already underline the appropriateness of the GGE approximation, we do not use more resources in order to propagate the dynamics all the way to the steady state.}

\end{appendix}

\bibliography{iterative_iris.bib}

\end{document}